# The Topology of Biological Networks from a Complexity Perspective

Ali Atiia
atiia@cs.mcgill.ca

François Major
major@iro.umontreal.ca

Jérôme Waldispühl
jeromew@cs.mcgill.ca

## *Abstract*

A complexity-theoretic approach to studying biological networks is proposed. A simple graph representation is used where molecules (DNA, RNA, proteins and chemicals) are vertices and relations between them are directed and signed (promotional (+) or inhibitory (-)) edges. Based on this model, the network evolution problem (NEP) is defined formally as an optimization problem and subsequently proven to be fundamentally hard (NP-hard) by means of reduction from the Knapsack problem (KP). For empirical validation, various biological networks of experimentally-validated interactions are compared against randomly generated networks with varying degree distributions. An NEP instance is created using a given real or synthetic (random) network. After being reverse-reduced to a KP instance, each NEP instance is fed to a KP solver and the average achieved knapsack value-to-weight ratio is recorded from multiple rounds of simulated evolutionary pressure. The results show that biological networks (and synthetic networks of similar degree distribution) achieve the highest ratios at maximal evolutionary pressure and minimal error tolerance conditions. The more distant (in degree distribution) a synthetic network is from biological networks the lower its achieved ratio. The results shed light on how computational intractability has shaped the evolution of biological networks into their current topology.

## 1. *Introduction*

At the intersection between biology and computer science lie two related areas of scientific inquiry. In one direction, *in silico* models of biological components/processes are used to gain insight into biology through algorithmics/statistics as exemplified in the field of bioinformatics. On the opposite direction, and since Adleman's insightful paper [1], computational problems (algorithms) are modeled (implemented) using biological components (processes) in a line of research that has rapidly evolved into the field of molecular computing. The computability/complexity [2] theory has not, however, been applied for the decipherment of actual biological systems. Given that the field of systems biology still lacks a foundational theory that can enable rigorous treatment of biological systems in a holistic manner, the complexity theory of computer science is proposed here for that purpose. It is a theory about machines, and biological systems are considered as such. Given its epistemological independence of physics and statistics, complexity theory can provide new insights into the evolution and functioning of biological networks, especially that "the role that natural selection has in the evolution of network structure remains unknown [3]."

In this work, a simple graph representation is used whereby nodes represent molecules and directed weighted edges represent promotional (positive weight) or inhibitory (negative weight) interactions. The addressed question is: given the universality (machine-independence) of the computability/complexity law, how do the effects of computational intractability manifest themselves in the evolution of biological networks? The network evolution problem (NEP) is defined formally and proved to be fundamentally hard (NP-hard) by reduction from the Knapsack Problem (KP).

For empirical demonstration, biological networks of experimentally validated interactions are compared to randomly generated networks with varying degrees of connectivity. Networks serve as KP instances (obtained by reverse-reduction) fed to a KP solver as input, and the latter's total knapsack value and weight is recorded for each network from multiple rounds of evolutionary pressure simulated by a hypothetical Oracle advice. The results show a clear link between a network's distance from biological networks (in degree distribution) and the maximum knapsack value-to-weight ratio achieved. In other words, biological networks show better adaptability as evolutionary pressure increases, and tolerance for errors decreases.

In Section 2, NEP is defined formally (2.1) and described informally (2.2). In section 3, theoretical and empirical results are presented. First, the NP-hardness of NEP is proven by reduction from the KP (3.1). Second, the data used in the empirical study is described (3.2). Third, the algorithmic workflow (3.3) and empirical results of computer simulations over the data are presented (3.4), comparing biological and synthetic networks. In section 4, conclusion a discussion and reflection on the results are presented.

## 2. *Method*

The evolution of biological networks is represented as a computational optimization problem. The definition is intended to compromise between opposing constraints: generality, to avoid symbolic bloat and artificial complexity, and specificity, to capture the reality of biological systems. It is also crucial that the proposed definition of the problem and the theoretical analysis of its complexity are easily falsifiable through empirical evidence. In this section, the network evolution (NEP) problem is defined formally then described informally, highlighting the correspondence between the model and actual biological systems.

*2.1 Formal Definition of the Network Evolution Problem (NEP):*

**Given:**

$$\boldsymbol{G} = \{g_1, g_2, \ldots, g_n\}$$

$$\boldsymbol{A} = a_1 a_2 \ldots \ldots a_n \qquad \text{where} \quad a_j \in \{0, -1, +1\}$$

$$\boldsymbol{P} = \sum_{i=1}^{n} |a_i|$$

$$\boldsymbol{M} = \begin{bmatrix} I_{11} & I_{12} & \cdots & I_{1n} \\ I_{21} & I_{22} & & I_{2n} \\ \vdots & & \vdots & \vdots \\ \vdots & & \vdots & \vdots \\ I_{n1} & I_{n2} & \cdots & I_{nn} \end{bmatrix} \quad \text{where} \quad I_{jk} \in \mathbb{R}$$

$$\boldsymbol{T} \in \mathbb{R}^{+}$$

**Let:**

$$\boldsymbol{B} = \{b_1, b_2, \ldots, b_n\} \qquad \text{where:}$$

$$b_j = \sum_{k=1}^{n} I_{jk} \oplus a_k \qquad \text{and} \qquad I_{jk} \oplus a_k = \begin{cases} |I_{jk}| & \text{if} \quad I_{jk} \times a_k > 0 \\ 0 & \text{otherwise} \end{cases}$$

$$\boldsymbol{D} = \{d_1, d_2, \ldots, d_n\} \qquad \text{where:}$$

$$d_j = \sum_{k=1}^{n} I_{jk} \ominus a_k \qquad \text{and} \qquad I_{jk} \ominus a_k = \begin{cases} |I_{jk}| & \text{if} \quad I_{jk} \times a_k < 0 \\ 0 & \text{otherwise} \end{cases}$$

Define $\boldsymbol{f}: \boldsymbol{G} \to \{0,1\}$ that:

$$maximizes \sum_{j=1}^{n} b_j \cdot f(g_j) \quad subject\ to \quad \left( \sum_{j=1}^{n} d_j \cdot f(g_j) \right) \leq T$$

## *2.2 Informal Description of the Network Evolution Problem (NEP):*

| | |
|---|---|
| $\boldsymbol{G} = \{g_1, g_2, \ldots, g_n\}$ | A set of ***G***enes: any transcribable element on the genome. |
| $\boldsymbol{A} = a_1 a_1 \ldots a_n \quad a_j \in \{0, -1, +1\}$ | A ternary string representing an Oracle ***A***dvice: $a_j = \begin{cases} 0 & \Rightarrow \quad \textit{no advice on } g_j \\ -1 & \Rightarrow \quad g_j \ \textit{should be repressed} \\ +1 & \Rightarrow \quad g_j \ \textit{should be promoted} \end{cases}$ |
| $\boldsymbol{P} = \sum_{i=1}^{n} \lvert a_i \rvert$ | ***P*** = ***p***ressure is the number of nodes towards which the Oracle is *not* indifferent. |
| $\boldsymbol{M} = \begin{bmatrix} I_{11} & I_{12} & \cdots & I_{1n} \\ I_{21} & I_{22} & & I_{2n} \\ \vdots & & \vdots & \vdots \\ \vdots & & \vdots & \vdots \\ I_{n1} & I_{n2} & \cdots & I_{nn} \end{bmatrix}$ | An ***I***nteraction ***M***atrix: $I_{jk} = \begin{cases} 0 & \Rightarrow \quad g_j \quad \textit{neutral to} \quad g_k \\ \alpha \in \mathbb{R}^+ & \Rightarrow \quad g_j \quad \textit{promotes} \quad g_k \\ \alpha \in \mathbb{R}^- & \Rightarrow \quad g_j \quad \textit{represses} \quad g_k \end{cases}$ |
| $\boldsymbol{T} \in \mathbb{R}^+$ | ***T*** = ***T***olerance is a threshold on how much total damage $\left(\sum_{j=1}^{n} d_j \cdot f(g_j)\right)$ is to be tolerated |
| $\boldsymbol{B} = \{b_1, b_2 \ldots, b_n\}$<br>$\boldsymbol{D} = \{d_1, d_2, \ldots, d_n\}$ | Sets of ***B***enefits and ***D***amages; each gene $g_i$ has a corresponding ***b***enefit value $\boldsymbol{b_i}$ and ***d***amage value $\boldsymbol{d_i}$. |
| $I_{jk} \oplus a_k = \begin{cases} \lvert I_{jk} \rvert & \text{if } I_{jk} \times a_k > 0 \\ 0 & \text{otherwise} \end{cases}$ | If the effect of $\boldsymbol{g_j}$ on $\boldsymbol{g_k}$ is in agreement with what the oracle says $\boldsymbol{g_k}$ should be (i.e. $\boldsymbol{I_{jk}}$ and $\boldsymbol{a_k}$ have the same sign), then increment $\boldsymbol{b_j}$ (the benefit value $\boldsymbol{g_j}$) by $\lvert I_{jk} \rvert$. |
| $I_{jk} \ominus a_k = \begin{cases} \lvert I_{jk} \rvert & \text{if } I_{jk} \times a_k < 0 \\ 0 & \text{otherwise} \end{cases}$ | If the effect of $\boldsymbol{g_j}$ on $\boldsymbol{g_k}$ is in disagreement with what the oracle says $\boldsymbol{g_k}$ should be (i.e. $\boldsymbol{I_{jk}}$ and $\boldsymbol{a_k}$ have different signs), then increment $\boldsymbol{d_j}$ (the benefit value $\boldsymbol{g_j}$) by $\lvert I_{jk} \rvert$. |
| $\boldsymbol{f} : \boldsymbol{G} \rightarrow \{0,1\}$ | A function defining what the actual state of a gene is: forcibly promoted (1) if it's not already, or forcibly repressed (0) if it's not already. While the sequence $(a_1 a_1 \ldots \ldots a_n)$ describes ***what should be,*** the sequence $(f(g_1) f(g_2) \ldots \ldots f(g_n))$ describes ***what is***. |
| Define $\boldsymbol{f}: \boldsymbol{G} \rightarrow \{0,1\}$ that:<br>*Maximizes*: $\sum_{j=1}^{n} b_j \cdot f(g_j)$<br>*subject to*: $\left(\sum_{j=1}^{n} d_j \cdot f(g_j)\right) \leq T$ | The oracle advice can be imposed by forcibly repressing every gene $\boldsymbol{g_j}$ where $\boldsymbol{a_j} = -1$ and forcibly promoting every gene $\boldsymbol{g_k}$ where $\boldsymbol{a_k} = +1$. However:<br>➢ **Repressing** $g_j$ can inadvertently contribute to a violation of the oracle advice because $g_j$ is a also a **promoter (repressor)** of some $g_i$ that should in fact be **promoted (repressed)**; and<br>➢ **Promoting** $g_k$ can inadvertently contribute to a violation of the oracle advice because $g_k$ is also a **promoter (repressor)** of some $g_i$ that should in fact be **repressed (promoted).**<br>What subset of genes should forcibly be promoted/repressed (define *f*) such that the oracle's advice is as satisfied as possible (*maximize ... subject to...*)?<br>**The idealistic enforcement of an oracle advice is complicated by the reality of network connectivity.** |

# 3. *Results*

## 3.1 *The NP-hardness of NEP:*

### 3.1.1 *Definition of the* $\mathrm{KNAPSACK_{OPT}}$ *optimization problem:*

Given a set of objects $\boldsymbol{O} = \{o_1, o_2, \dots, o_r\}$

a set of values $\boldsymbol{V} = \{v_1, v_2, \dots, v_r\}$ $\quad v_i \in \mathbb{Z}^+$

a set of weights $\boldsymbol{W} = \{w_1, w_2, \dots, w_r\}$ $\quad w_i \in \mathbb{Z}^+$

a knapsack capacity $\boldsymbol{C} \in \mathbb{Z}^+$

Define:

$$\mathrm{KNAPSACK_{OPT}} = Maximize \sum_{i=1}^{r} x_i v_i \quad x_i \in \{0,1\}, \quad subject\ to \sum_{i=1}^{r} x_i w_i \leq C$$

$\mathrm{KNAPSACK_{OPT}}$ is NP-hard [4].

### 3.1.2 *Reduction of* $\mathrm{KNAPSACK_{OPT}}$ *to NEP:*

Given $\mathrm{KNAPSACK_{OPT}}$ definition above, define an NEP instance as follows:

$$\boldsymbol{G} = \{g_1, g_2, \dots, g_r\}$$

$$\boldsymbol{A} = a_1 a_2 \dots \dots a_r \qquad \text{where} \quad \forall j, 1 \leq j \leq r,\ a_j = +1$$

$$\boldsymbol{P} = r$$

$$\boldsymbol{M} = \begin{bmatrix} I_{11} & I_{12} & \cdots & I_{1r} \\ I_{21} & I_{22} & & I_{2r} \\ \vdots & & \vdots & \vdots \\ \vdots & & \vdots & \vdots \\ I_{r1} & I_{r2} & \cdots & I_{rr} \end{bmatrix} \quad \text{where} \quad I_{jk} = \begin{cases} \dfrac{v_j}{\lfloor r/2 \rfloor} & if \quad 1 \leq k \leq \left\lfloor \dfrac{r}{2} \right\rfloor \\ \\ \dfrac{-w_j}{\lceil r/2 \rceil} & if \quad \left\lfloor \dfrac{r}{2} \right\rfloor < k \leq r \end{cases}$$

$$\boldsymbol{T} = C$$

Follow NEP definition in section 2.1 and*:*

1. Calculate $\boldsymbol{B}$ and $\boldsymbol{D}$
2. Define $\boldsymbol{f}$
3. Return $\{x_1, x_2, \dots, x_r\}$ where $x_i = f(g_i)$

## *3.2 Data:*

A total of seven networks, three biological and four synthetic, are used in this comparative study. All biological networks are from experimentally-validated interactions, while synthetic networks are computer-generated using various random graph generating algorithms. Table 1 summarizes properties of each network. The first biological network is obtained from IntAct database (actively humanly curated) [5] by submitting a Molecular Interaction Query Language (MIQL) query through PSICQUIC [6] (Proteomics Standard Initiative Common QUery of InteraCtions) implementation[1]. The resulting interactions from the query meet the following constraints:

1. Homo sapiens endogenous molecules only for both interactors.
2. Direct interactions (direct physical contact).
3. Supported by experimental evidence (as opposed to those inferred manually or computationally).

The resulting network is comprised of 1779 nodes and 3272 interactions (see Appendix A.1 for further details). Since the direction and sign of interactions are not provided by IntAct database (or any other molecular interaction database to our knowledge), they are assigned randomly. The second and third biological networks used in this study, obtained from Surantee *et. al.* (Table S3 in [7]) and Vinayagam *et. al.* (Table S7 in [8]), respectively, do however assign direction and sign (inhibitory or promotional) to each interaction[2].

Synthetic network are obtained using various random graph-generating methods. The Scale-free and Barabasi-Albert (BA) networks have an average degree distribution (number of in- and out- edges) that follows a power law, making them similar to biological networks in network connectivity (last column in Table 1). Generally, in these two models, the probability of adding an edge to and/or from a node is proportional to its current degree (the more edges a node already has the more likely it will be further connected to new nodes). In the Erdos-Renyi (ER) model of generating random graphs [9], an edge between each pair of nodes is added with equal probability and independently of the current degree of both nodes. Lastly, Complete graph is one where each node is connected to every other node but itself. In all synthetic networks, both direction and sign of interactions (edges) are assigned randomly.

---

[1] http://www.ebi.ac.uk/Tools/webservices/psicquic/view/main.xhtml

[2] To our knowledge, these are the only two available datasets of experimentally validated, directed, and signed molecular interactions.

| | Network | Source | Edge Direction | Edge Sign | No. of nodes | Avg. no. of neighbors |
|---|---|---|---|---|---|---|
| Biological | IntAct | MIQL Query | random | random | 1779 | 3.3 |
| | Suratanee | Surantee *et.al.* [7] | inferred | inferred | 921 | 4.2 |
| | Vinayagam | Vinayagam *et.al.* [8] | inferred | inferred | 3058 | 3.8 |
| Synthetic | Scale-free | Computer-generated [10] | random | random | 2000 | 3.2 |
| | Barabási–Albert (BA) | Computer-generated [10][11] | random | random | 2000 | 6.0 |
| | Erdős–Rényi (ER) | Computer-generated [10][9] | random | random | 2000 | 399 |
| | Complete | Computer-generated [10] | random | random | 2000 | 1999 |

**Table 1**: **Summary of network data sets used in this study.** Random edge (interaction) direction and signs are assigned with a fair coin flip. Inferred edge are obtained computationally through a trained machine learning algorithm and its prediction accuracy is verified by cross-validation with known experimentally-validated directed/signed interactions (see text for details).

## *3.3 Algorithmic Workflow of Computer Simulation:*

The simulation has two parameters: evolutionary pressure $\boldsymbol{p}$ and tolerance $\boldsymbol{t}$. Evolutionary pressure is captured in the NEP definition by the Oracle advice $A$, and in each simulation we refer to $p = \sum_{i=1}^{n} |a_i|$ as the number of nodes towards which the Oracle is *not* indifferent. For example, for a network $G$ of $n$ nodes, $p$ =5 implies there is a subset of 5 nodes $S = \{n_i, n_j, n_k, n_l, n_m\} \subset G$ where $\forall n_x \in S, a_x \neq 0$ and $\forall n_y \notin S, a_y = 0$. The tolerance $T$ is the optimization threshold in NEP definition (Section 2.1) or, equivalently, the knapsack capacity $C$ in the corresponding KP instance (see reduction, section 3.1). For each network described in Section 3.2, a simulation is carried out for each $p = 5, 10, 50, \text{or } 500$ against each $t = 5, 10, 50, \text{or } 500$.

Given a $(p, t)$ pair, a knapsack instance is generated from a given network by reversing the reduction shown Section 3.1.2, that is: $O = G$, $V = B$, $W = D$, and $C = T$. The simulation records the total value and weight of objects (=nodes, recall $O = G$) added to the knapsack by the solver ($\sum x_i v_i \quad x_i \in \{0,1\}$) for each round against a randomly generated Oracle advice on $p$ nodes. The simulation is repeated $i$=10,000 times (sampling threshold). Simulation time runs exponentially (days of execution time on 128 parallel CPUs). Using $i$=100,000 on IntAct and Suratanee networks over days of simulations produced virtually the same results (Appendix A.2). In the case of the Complete network, one round of simulation is sufficient (because its nodes have the same exact degree and edges have equal probability of direction/sign values). Figure 1 below summarizes the algorithmic workflow of the simulation.

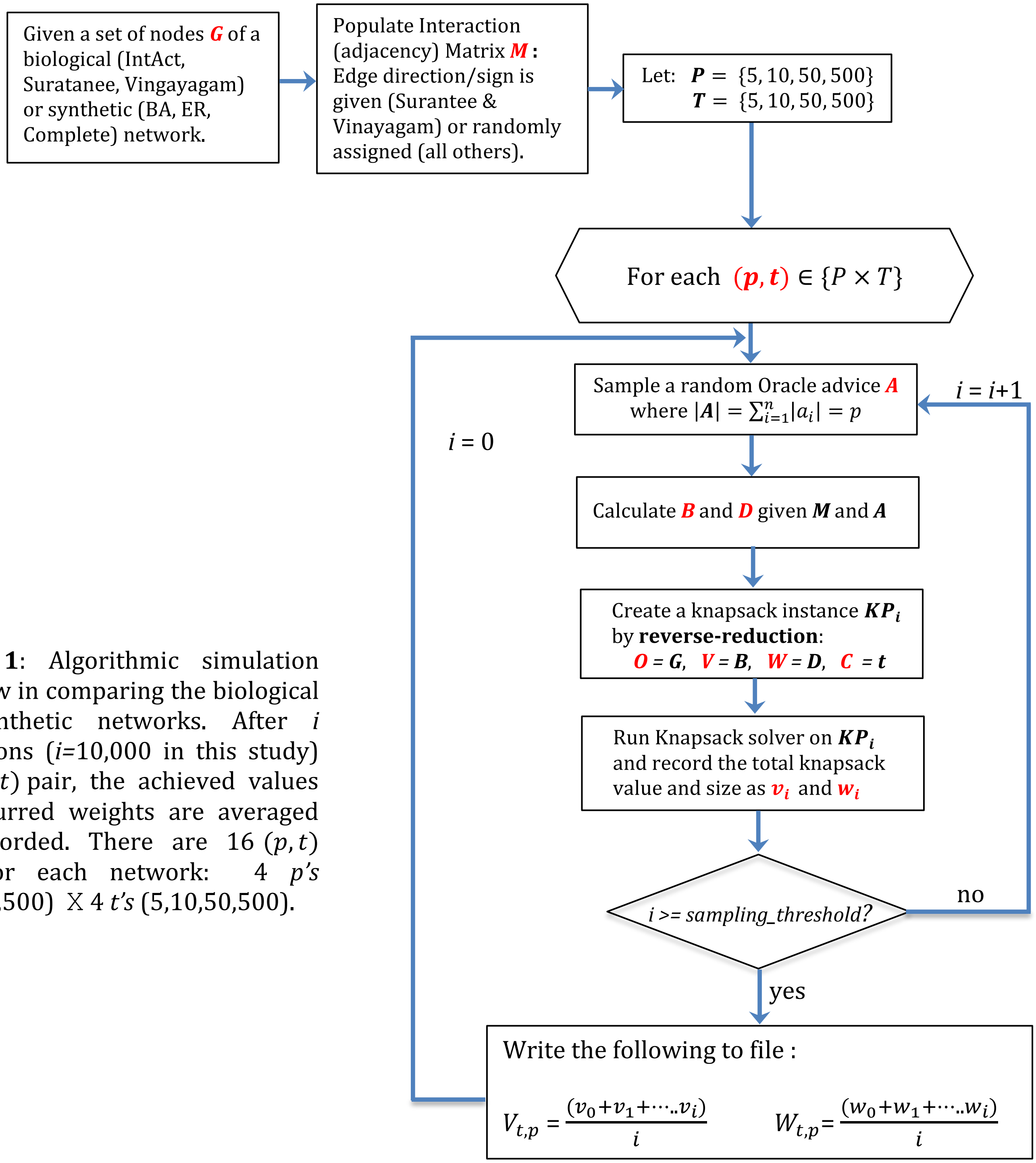

**Figure 1**: Algorithmic simulation workflow in comparing the biological and synthetic networks. After *i* simulations (*i*=10,000 in this study) on a $(p, t)$ pair, the achieved values and incurred weights are averaged and recorded. There are 16 $(p, t)$ pairs for each network: 4 *p's* (5,10,50,500) X 4 *t's* (5,10,50,500).

### *3.4 Simulation Results:*

Table 2 and 3 show the average V and W values achieved after 10k rounds of simulation on each $(p, t)$ pair for each network (except for the Complete network where 1 round for each $(p, t)$ pair is sufficient). The highest V/W ratio achieved is highlighted (orange), and it shows a clear superior adaptability in biological (IntAct, Suratee, and Vingayam) and pseudo-Biological (BA and Scale-free) networks as they achieve highest scores at maximum pressure $p$ and lowest tolerance $t$. Networks that are very distant in degree distributions from biological networks (ER and Complete) achieve high scores at low pressure, but display poor results as the pressure increases. Nonetheless, biological and pseudo-biological networks also show lower V/W ratio when tolerance increases but constant pressure (compare constant $p$ = 500 to variable $t$ =5, 10, 50 and 500). This may seem paradoxical since, conceptually, increased tolerance should imply improved adaptability. However, observing the same data (constant $p$, variable $t$) shows that increased tolerance affects V/W ration not by lowering V but rather by increasing W, as seen in Figure 2.

| | | Biological Networks | | | | | | | | |
|---|---|---|---|---|---|---|---|---|---|---|
| | | IntAct | | | Suratee | | | Vingayam | | |
| | | n=1779 | | | n=967 | | | n=3058 | | |
| $p$ | $t$ | V | W | V/W | V | W | V/W | V | W | V/W |
| 5 | 5 | 4.0 | 0.0 | 4.0 | 5.0 | 0.0 | 5.0 | 4.0 | 0.0 | 4.0 |
| | 10 | 4.0 | 0.0 | 4.0 | 4.0 | 0.0 | 4.0 | 4.0 | 0.0 | 4.0 |
| | 50 | 4.0 | 0.0 | 4.0 | 4.0 | 0.0 | 4.0 | 4.0 | 0.0 | 4.0 |
| | 500 | 4.0 | 0.0 | 4.0 | 5.0 | 0.0 | 5.0 | 4.0 | 0.0 | 4.0 |
| 10 | 5 | 9.0 | 0.0 | 9.0 | 10.0 | 0.0 | 10.0 | 9.0 | 0.0 | 9.0 |
| | 10 | 9.0 | 0.0 | 9.0 | 10.0 | 0.0 | 10.0 | 9.0 | 0.0 | 9.0 |
| | 50 | 9.0 | 0.0 | 9.0 | 10.0 | 0.0 | 10.0 | 9.0 | 0.0 | 9.0 |
| | 500 | 9.0 | 0.0 | 9.0 | 10.0 | 0.0 | 10.0 | 9.0 | 0.0 | 9.0 |
| 50 | 5 | 43.0 | 3.0 | 14.3 | 48.0 | 3.0 | 16.0 | 45.0 | 1.0 | 45.0 |
| | 10 | 45.0 | 6.0 | 7.5 | 50.0 | 5.0 | 10.0 | 45.0 | 1.0 | 45.0 |
| | 50 | 46.0 | 7.0 | 6.6 | 50.0 | 5.0 | 10.0 | 45.0 | 1.0 | 45.0 |
| | 500 | 45.0 | 7.0 | 6.4 | 50.0 | 5.0 | 10.0 | 45.0 | 1.0 | 45.0 |
| 500 | 5 | 268.0 | 4.0 | 67.0 | 241.0 | 4.0 | 60.3 | 334.0 | 4.0 | 83.5 |
| | 10 | 278.0 | 9.0 | 30.9 | 255.0 | 9.0 | 28.3 | 346.0 | 9.0 | 38.4 |
| | 50 | 340.0 | 49.0 | 6.9 | 334.0 | 49.0 | 6.8 | 397.0 | 49.0 | 8.1 |
| | 500 | 458.0 | 206.0 | 2.2 | 505.0 | 281.0 | 1.8 | 454.0 | 132.0 | 3.4 |

**Table 2: Results of simulation on biological networks.** The V, W are average values of 10k round of simulations on a given $(p, t)$ pair. The V/W column shows the ratio of value-to-weight achieved. Maximum ratios for each network are highlighted (orange); n=number of nodes in a network.

| | | Synthetic (computer-generated) Networks | | | | | | | | | | | |
|---|---|---|---|---|---|---|---|---|---|---|---|---|---|
| | | BA | | | Scale-free | | | ER | | | Complete | | |
| | | n=2000 | | | n=2000 | | | n=2000 | | | n=2000 | | |
| $p$ | $t$ | V | W | V/W | V | W | V/W | V | W | V/W | V | W | V/W |
| 5 | 5 | 7.0 | 0.0 | 7.0 | 3.0 | 0.0 | 3.0 | 416.0 | 4.0 | 104.0 | 761.0 | 5.0 | 152.2 |
| | 10 | 7.0 | 0.0 | 7.0 | 3.0 | 0.0 | 3.0 | 422.0 | 9.0 | 46.9 | 576.0 | 9.0 | 64.0 |
| | 50 | 7.0 | 0.0 | 7.0 | 3.0 | 0.0 | 3.0 | 462.0 | 49.0 | 9.4 | 740.0 | 50.0 | 14.8 |
| | 500 | 7.0 | 0.0 | 7.0 | 3.0 | 0.0 | 3.0 | 499.0 | 92.0 | 5.4 | 1734.0 | 499.0 | 3.5 |
| 10 | 5 | 14.0 | 0.0 | 14.0 | 7.0 | 0.0 | 7.0 | 641.0 | 4.0 | 160.3 | 287.0 | 5.0 | 57.4 |
| | 10 | 14.0 | 0.0 | 14.0 | 7.0 | 0.0 | 7.0 | 655.0 | 9.0 | 72.8 | 429.0 | 10.0 | 42.9 |
| | 50 | 14.0 | 0.0 | 14.0 | 7.0 | 0.0 | 7.0 | 733.0 | 49.0 | 15.0 | 591.0 | 50.0 | 11.8 |
| | 500 | 14.0 | 0.0 | 14.0 | 7.0 | 0.0 | 7.0 | 998.0 | 368.0 | 2.7 | 1773.0 | 499.0 | 3.6 |
| 50 | 5 | 67.0 | 4.0 | 16.8 | 41.0 | 0.0 | 41.0 | 437.0 | 4.0 | 109.3 | 130.0 | 4.0 | 32.5 |
| | 10 | 71.0 | 8.0 | 8.9 | 39.0 | 1.0 | 39.0 | 470.0 | 9.0 | 52.2 | 182.0 | 9.0 | 20.2 |
| | 50 | 74.0 | 13.0 | 5.7 | 39.0 | 1.0 | 39.0 | 672.0 | 49.0 | 13.7 | 272.0 | 50.0 | 5.4 |
| | 500 | 74.0 | 13.0 | 5.7 | 40.0 | 1.0 | 40.0 | 1864.0 | 499.0 | 3.7 | 1321.0 | 500.0 | 2.6 |
| 500 | 5 | 296.0 | 4.0 | 74.0 | 313.0 | 4.0 | 78.3 | 0.0 | 0.0 | 0.0 | 23.0 | 5.0 | 4.6 |
| | 10 | 314.0 | 9.0 | 34.9 | 325.0 | 9.0 | 36.1 | 15.0 | 5.0 | 3.0 | 52.0 | 10.0 | 5.2 |
| | 50 | 403.0 | 49.0 | 8.2 | 365.0 | 46.0 | 7.9 | 128.0 | 49.0 | 2.6 | 126.0 | 50.0 | 2.5 |
| | 500 | 748.0 | 471.0 | 1.6 | 398.0 | 92.0 | 4.3 | 1050.0 | 500.0 | 2.1 | 878.0 | 500.0 | 1.8 |

**Table 3: Results of simulation results on synthetic networks.** The V, W are average values of 10k round of simulations on a given $(p, t)$ pair. The V/W column shows the ratio of value-to-weight achieved. Maximum ratios for each network are highlighted (orange); n=number of nodes in a network. BA and Scale-free networks, which are of similar degree distributions to biological networks [11], expectedly show similar results to those in Table 2. ER and Complete networks show poor adaptability as $p$ increases, even with relaxed tolerance threshold $t$.

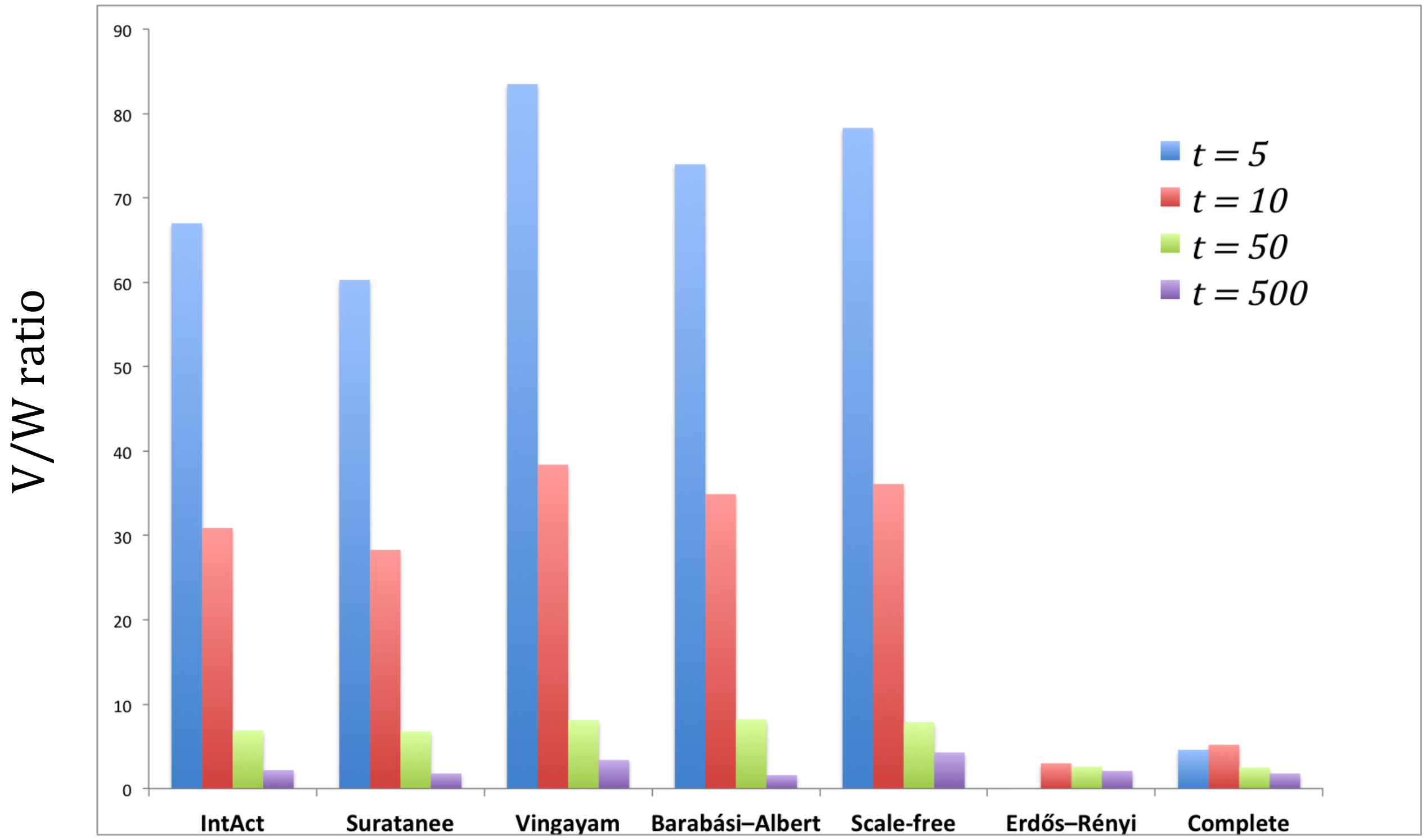

**Figure 2: Network adaptability under maximum evolutionary.** The graph shows the V/W ratio at the highest simulated evolutionary pressure $p$ = 500 and increasing degrees of tolerance $t = 5, 10, 50,$ and 500. Blue columns represent maximum pressure ($p = 500$) at minimal tolerance ($p = 5$). Biological networks (IntAct, Suratanee and Vingayam) as well as synthetic networks BA and Scale-free (which are similar in degree distributions to biological networks [11]) achieve higher V/W ratio under maximum evolutionary pressure and minimal tolerance.

## 4. *Discussion*

This study is an attempt towards linking systems biology to the complexity theory of computer science. The aim was to employ the theoretical for the sake of the practical, and so the presented definition of the network evolution problem was intended to be general enough to avoid symbolic bloat and artificial complexity (which hinders falsifiability) but also specific enough to capture the intricacies of biological systems, ensuring its relevance and utility in the practical pursuit of understanding actual biological systems holistically. The empirical study juxtaposes biological networks to various synthetic ones, and the presented simulation results show clear indications that computational intractability does shape the evolution of biological networks (in their degree distributions, particularly). Further validation using a more systematic and comprehensive survey of synthetic networks is underway.

# Appendix

## A.1 *The IntAct network*:

The syntax of MILQ query against IntAct database is (9606 = *Homo sapiens*):

```
taxidA:9606 AND  taxidB:9606 AND
type:"direct interaction"    AND
detmethod:"experimental interaction detection"
```

The query imposes the constraints described in Section 2.3. The query returns a molecular interaction network of 2238 nodes and 5328 interactions (edges), which are further filtered down in Cytoscape [8] by removing small islands, duplicate edges, and self-loops, leaving a total of 1779 nodes and 3272 interactions, depicted graphically in Figure A.1 below.

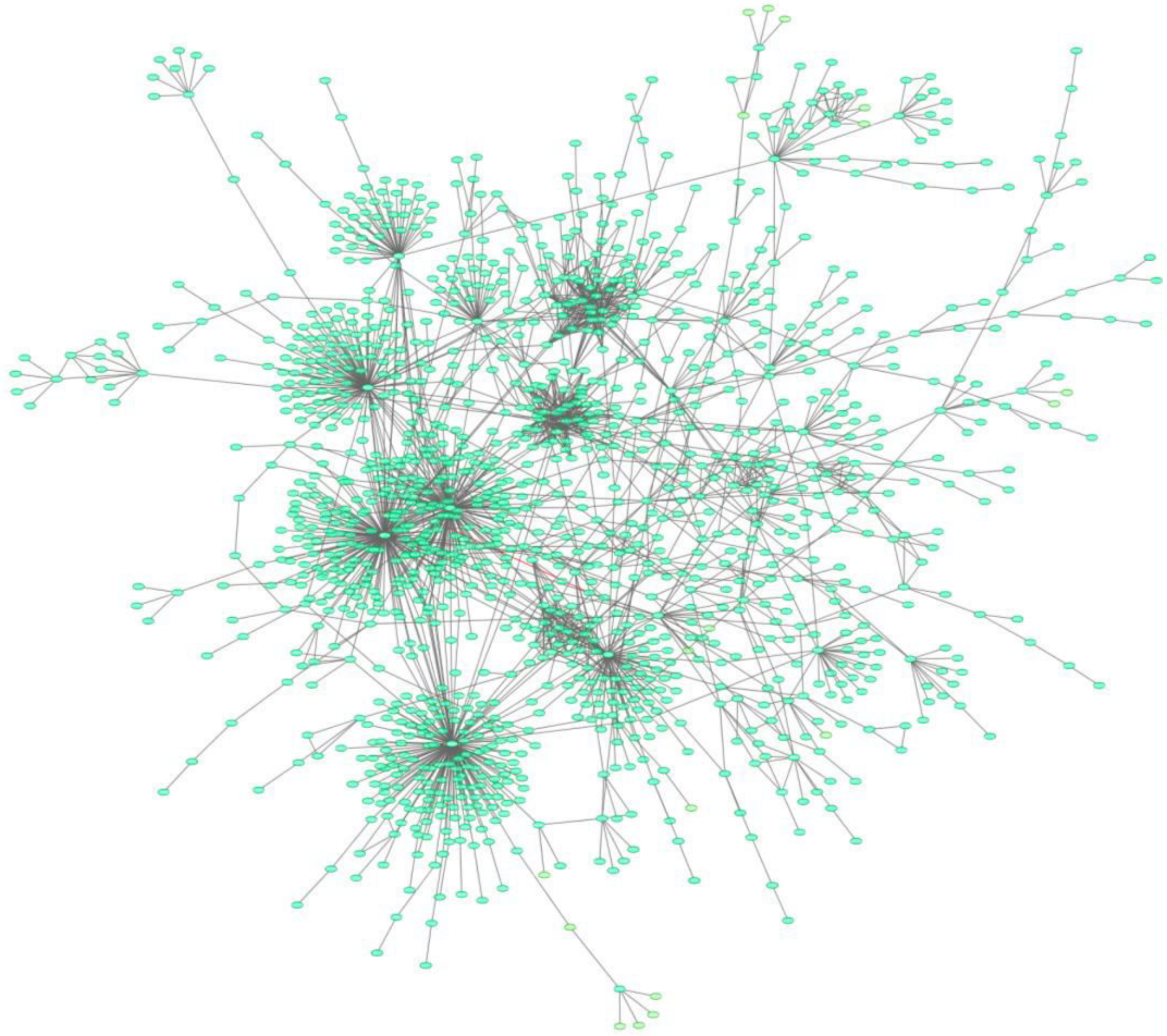

**Figure A.1:** The resulting network of directly interacting molecules in *homo sapiens*-only samples with supporting experimental evidence, obtained from IntAct database (see text, section 3.2). Excluded from the network are self-loops and small islands. There are a total of 3272 interactions (edges) and 1779 interactors (nodes).

**A.2** *Simulation threshold:*

Increasing the sampling threshold in each round of (p,t) pair to *i*=100,000 (i.e. 100k rounds of selction-without-replacement of *k* nodes to be presented for an Oracle advice) produce virtually the same results as *i*=10,000. It is worth noting that each V and W value presented is the average of all 10,000 or 100,000 V's recorded in each round of (p,t) pair.

| | | IntAct Network | | | | | |
|---|---|---|---|---|---|---|---|
| | | 10k simulations on each $(p,t)$ | | | 100k simulations on each $(p,t)$ | | |
| | | n=1779 | | | n=1779 | | |
| $p$ | $t$ | V | W | V/W | V | W | V/W |
| 5 | 5 | 1.00 | 0.00 | 1.00 | 1.00 | 0.00 | 1.00 |
| | 10 | 1.00 | 0.00 | 1.00 | 1.00 | 0.00 | 1.00 |
| | 50 | 1.00 | 0.00 | 1.00 | 1.00 | 0.00 | 1.00 |
| | 500 | 1.00 | 0.00 | 1.00 | 1.00 | 0.00 | 1.00 |
| 10 | 5 | 3.00 | 0.00 | 3.00 | 3.00 | 0.00 | 1.00 |
| | 10 | 3.00 | 0.00 | 3.00 | 3.00 | 0.00 | 1.00 |
| | 50 | 3.00 | 0.00 | 3.00 | 3.00 | 0.00 | 1.00 |
| | 500 | 3.00 | 0.00 | 3.00 | 3.00 | 0.00 | 1.00 |
| 50 | 5 | 17.00 | 1.00 | 17.00 | 17.00 | 1.00 | 17.00 |
| | 10 | 17.00 | 1.00 | 17.00 | 17.00 | 1.00 | 17.00 |
| | 50 | 17.00 | 1.00 | 17.00 | 17.00 | 1.00 | 17.00 |
| | 500 | 17.00 | 1.00 | 17.00 | 17.00 | 1.00 | 17.00 |
| 500 | 5 | 130.00 | 4.00 | 32.50 | 130.00 | 4.00 | 32.50 |
| | 10 | 138.00 | 9.00 | 15.33 | 138.00 | 9.00 | 15.33 |
| | 50 | 173.00 | 48.00 | 3.60 | 173.00 | 48.00 | 3.60 |
| | 500 | 177.00 | 58.00 | 3.05 | 177.00 | 58.00 | 3.05 |